# When lowering temperature, the *in vivo* circadian clock in cyanobacteria follows and surpasses the *in vitro* protein clock trough the Hopf bifurcation

Mihalcescu I[a,*], Kaji H[b], Maruyama H[b], Giraud J[a], Van-Melle Gateau M[a], Houchmandzadeh B[a], Ito H[b]

*Univ. Grenoble Alpes, CNRS, LIPHY, F-38000 Grenoble, France*
*Faculty of Design, Kyushu University, Fukuoka, 815-8540, Japan*
[*]*To whom correspondence may be addressed: Irina.Mihalcescu@univ-grenoble-alpes.fr*

## Abstract

The *in vivo* circadian clock in single cyanobacteria is studied here by time-lapse fluorescence microscopy when the temperature is lowered below 25°C. We first disentangle the circadian clock behavior from the bacterial cold shock response by identifying a sequence of "death steps" based on cellular indicators. By analyzing only "alive" tracks, we show that the dynamic response of individual oscillatory tracks to a step-down temperature signal is described by a simple Stuart-Landau oscillator model. The same dynamical analysis applied to *in vitro* data (KaiC phosphorylation level following a temperature step-down) allows for extracting and comparing both clock's responses to a temperature step down. It appears, therefore, that both oscillators go through a similar supercritical Hopf bifurcation. Finally, to quantitatively describe the temperature dependence of the resulting *in vivo* and *in vitro* Stuart-Landau parameters $\mu(T)$ and $\omega_c(T)$, we propose two simplified analytical models: temperature-dependent positive feedback or time-delayed negative feedback that is temperature compensated. Our results provide strong constraints for future models and emphasize the importance of studying transitory regimes along temperature effects in circadian systems.

# 1. Introduction

Circadian rhythm is a self-sustaining biological oscillator evolved to adapt to Earth's day-night cycle. It has a ~24-hour period, can be synchronized by external cues, and is temperature compensated. These properties are conserved across species, aiding survival in diverse environments. However, when temperatures lowers, early studies in algae[1], and more recent ones in hibernating animals[2], in a population of cyanobacteria[3] and in neurons within cultured brain slices[4] have shown that the circadian rhythm disappears. Our focus here is on the dynamic process that leads to the cessation of the clock and the insights this may offer into current models of the circadian clock.

For this purpose, we focus our study on the circadian clock in the cyanobacterium *Synechococcus elongatus* PCC7942. The main properties of the circadian clock are derived from the core posttranscriptional oscillator[5], referred to here as the *in vitro* clock, which requires only three proteins (KaiA, KaiB, KaiC) and ATP to reconstitute a circadian rhythm. In living bacteria, this core oscillator, imbedded in multiple feedback loops related to transcription, translation, degradation, or uneven distribution of cellular components, operates as the *in vivo* oscillator. When bacteria are grown at the optimal growth temperature (30°C), the properties of the *in vitro* clock are enhanced demonstrating high temporal precision[6] and resistance to fluctuations[7].

Lowering the temperature reduces the oscillation amplitude of the *in vitro* oscillator and slightly affects its frequency[3]. Below a critical temperature ($T_c$) of approximately 19°C, the clock fades, indicating that a supercritical Hopf bifurcation occurs at this point. In the same work [3] it was found that no oscillations are observed in bacterial populations when the temperature drops below 20°C, although the mechanism for this disappearance remains unclear. Two scenarios are possible: cells may desynchronize while maintaining individual oscillations, or the *in vivo* oscillations of each cell may cease due to a Hopf bifurcation. A previous report [8], suggests the first hypothesis, as exposing single bacteria to varying temperature pulses (25°C-35°C) can stochastically disperse the phases of the oscillator, revealing an unstable singularity. However, with a permanent temperature drop, all parameters of the clock may change, potentially transforming the unstable singularity into an attractor and causing the clock's oscillatory pattern to vanish into an arrested state.

Her we investigate the behavior of *in vivo* oscillators at the single-cell level over an extended duration (7 days) and demonstrate that in fact the individual *in vivo* oscillators go through a Hopf bifurcation. We propose a simple model of interaction between the *in vitro* core oscillator and cellular machinery that captures the main features of the *in vivo* circadian clock under a permanent temperature drop. To reach these conclusions, we developed new experimental and analytical tools.

One challenge was the high mortality rate of bacteria exposed to cold temperatures during time-lapse fluorescence microscopy. We created a procedure to differentiate dying cell tracks from those of living cells, detailed in Section 2. We then focused on living cells, measuring the amplitude and period of each individual cell as the temperature shifted from 25 °C to colder ranges (21-16 °C). This analysis, outlined in Section 3, emphasizes the importance of the transitory phase, using the Stuart-Landau model to analyze the transitory signal and extract key parameters. Our findings confirm that individual oscillators experience a Hopf bifurcation. In Section 4, we integrate previously published data from the core oscillator with our current study, proposing a simple model for the feedback and interactions between these oscillators. Section 5 discusses our results and potential future developments.

## 2. Monitoring live bacteria at low temperatures

Cyanobacteria taxa are known to be widespread throughout the biosphere, often with persistent cold temperatures and freeze-thaw cycles, like polar or alpine environments [9]. Low temperature affects bacteria in terms of immediate decrease in membranes fluidity and in the efficiency of transcription and translation [10]. In response, bacteria act with a cold shock response to protect cellular functions, particularly photosynthesis, against the adverse effects of cold [11]. The strain studied here, *Synechococcus elongatus* (PCC7942), is a 40-year lab strain originated from California freshwater isolate [12], with maximal growth rate at 40°C [13, 14] and largely studied at 30°C. A slight decrease in temperature, to 25°C, already triggers a cold stress response (8) and incubation for 8 hours at 15°C is considered as severe non-permissive, resulting in massive cell death [16]. Therefore, to be able to follow the circadian clock at temperatures as low as 16°C-18°C, for time interval as long as a couple of periods (thus a few days in a row) bacteria must first survive and we must untangle cold stress response from circadian clock response.

It has been shown that pre-adaptation of cyanobacteria to intermediate low temperatures (25°C) increases survival when plunging to even lower non-acclimative temperatures [11]. For this purpose, all our liquid precultures were first grown at 25°C, and then installed in a the microscope observation chamber[17] while keeping the temperature at 25°C for 65-80h before lowering it to a range 15°C-18°C (Fig.1a). We used two strains, the wild-type PCC7942 and a deletion of the clock genes *(ΔkaiBC)*. Both strains carry a circadian activity fluorescence monitor, the tagged *yfp* gene (*yfp-lvA*) under the control of the P*kaiBC* promoter [17,18]. The kaiBC promoter drives oscillatory production of YFP in the wild-type strain, while in the Δ*kaiBC* strain performs as a constitutive promoter.

Figure 1a shows a film strip of the YFP fluorescence for colonies originating from single cells. For the first 65 hours, at 25°C, the bacteria are actively growing and dividing, regardless of whether they are wild type or Δ*kaiBC*. During this time, the YFP fluorescence monitor for the two wild-types (*wt*) sets of cells (*wt-1, wt-2*) oscillates with a 24-hour period (peaking at 0 and 24 h, and troughing at 12 and 36 h). In contrast, in the Δ*kaiBC* cells, the YFP fluorescence continuously increases, as expected for a non-oscillating strain.

Figures 1b,c,d depict respectively the instantaneous YFP fluorescence, red autofluorescence [17], and the length of these initial bacteria followed along one of their descendants lineage. The red fluorescence monitored in this experiment is related to the thylakoid membrane autofluorescence and is used as a complementary probe of cellular state. After 65 hours, the temperature is shifted to 16°C, and the cells growth significantly slows down, leading to a gradual emergence of two different scenarios in their outcome: in the first scenario, the cells exhibit slow growth all along the experiment and even occasionally divide, while in the second scenario, cell growth stops completely.

In Figure 1d, the green line denotes a bacterium that continues to grow at cold temperature: its YPF fluorescence oscillates (Fig.1b), and the fluorescence of the thylakoid membranes, varies only slightly (Fig.1c) without any major changes in either intensity or cellular localization. For all other tracks presented in Figure 1d, *i.e.* the two lineages descendants of *wt-2* colony (blue lines) and Δ*kaiBC* colony (grey lines)*,* the cold temperature introduces drastically different effects: after a significant decrease in YFP fluorescence lasting $\cong$ 24 h, a well-defined minimum is observed. This specific time coincides with growth arrest and even cell size deflation (see the corresponding-colored arrows in Fig. 1b and 1d). Approximately 24 h after this minimum, many events are observed: a peak in red fluorescence, a complete disorganization of thylakoid membranes (as shown in the inset red fluorescence images before and after the peak in Fig. 1c), and, in some cases, evident vacuolization (see Supplementary Information). This is a clear signature of cellular death, and we will next briefly inquire into the events before and after this occurrence.

As these last major changes do not occur synchronously, we first selected every track in the experiment where a strong peak in the red autofluorescence was detected. Figure 1e, f and g represent the YFP, red fluorescence and cell size, superimposed for each of those tracks: *wt* (green, red, magenta lines) and *ΔkaiBC* (grey lines). Next, we time-recentered all the tracks to their time of red fluorescence peak. A common trend emerges for both strains (*wt* or *ΔkaiBC*), regardless of the presence or absence of the circadian clock genes: normalized red fluorescence exhibits the same time dependence (Fig. 1i) and the YFP fluorescence clock reporter has a similar 72h oscillation (Fig. 1h). This demonstrates that this phenomenon is not clock related but is certainly associated with a sequence of events leading to cell death. Interestingly, if we normalize the cell size to the one 48 h after the red fluorescence peak, another common trend appears: all cell growth ceases no later than 48h before the red fluorescence peak (Fig. 1j, green ellipsis).

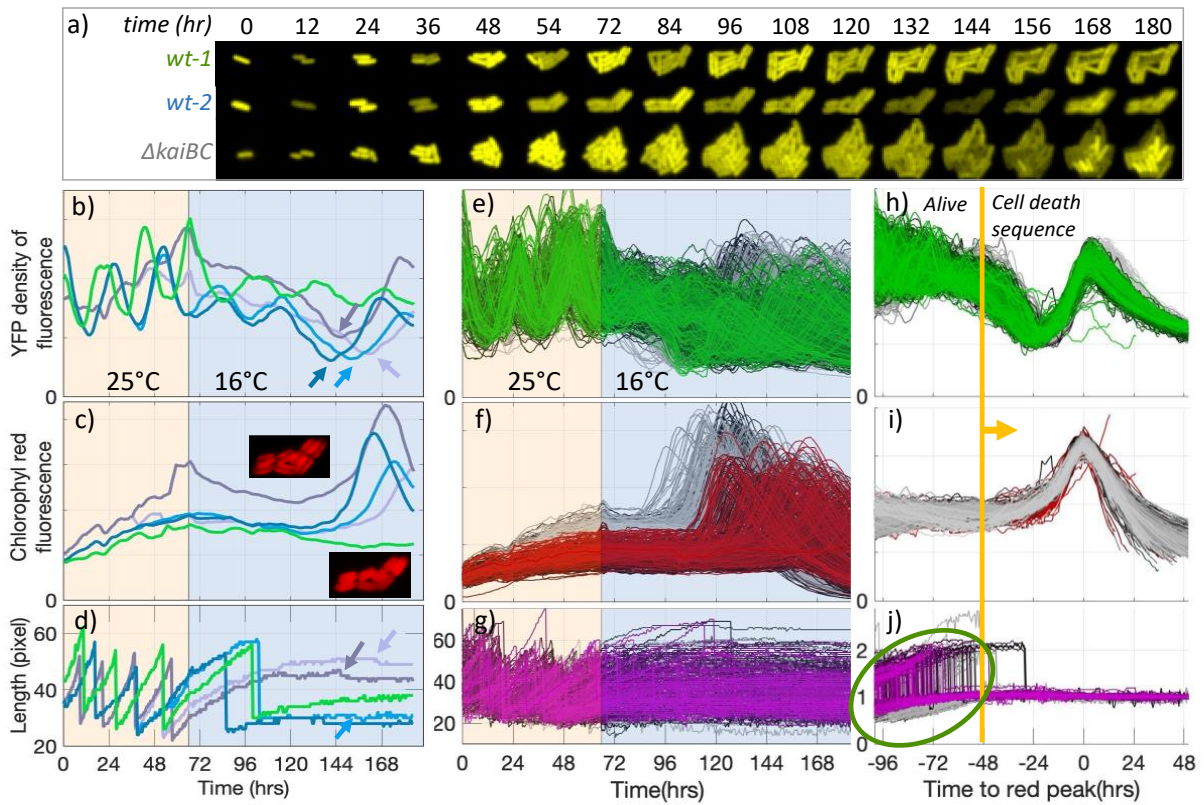

*Figure 1 Separate alive from dying tracks at low temperature a) Film strip of yfp fluorescence channel for 3 representative microcolonies as a function of time (in hours from the beginning of the experiment). The background color (yellow to blue) indicates the temperature (25°C or 16°C). The first two microcolonies are wild-type (wt-1 harbors surviving cells, while in wt-2 all bacteria are dead in the end) and the last one ΔkaiBC. b-d) The bacterial characteristics (b- YFP fluorescence density/cell, c- chlorophyll red fluorescence density and d- cell length in pixels) of five representative tracks chosen from the a) microcolonies as a function of time. The green track, a survivor example, is from the microcolony wt-1, two other tracks (in blue shades) from wt-2 and last two (grey shades) from ΔkaiBC. The insert in c) represent two snapshots of wt-2, on the red channel for the thylakoid membrane fluorescence, before and after the peak. The arrow pinpoint to the specific cell whose tracks are followed. e)-g) an overall view of all the tracks in this experiment harboring a peak in the red autofluorescence: e)- YFP fluorescence density/cell, respectively shades of green for wt and grays for ΔkaiBC f- chlorophyll red fluorescence density, shades of red for wt and grays for ΔkaiBC , and g- cell length in pixels), shades of purple for wt and grays for ΔkaiBC .h-j) the same tracks, once sorted by alignment by the time of the red fluorescence peak. The time is in hours relative to the peak time. The yfp channel (h) aligns well up to 48 hours prior to the peak (-48h, yellow vertical line), for both wt or ΔkaiBC tracks. From the same time on, all growth stops (j), the realigned cell length tracks are flat on the left side of the vertical yellow line.*

The orange vertical line in Figures 1h, i, and j marks a critical juncture. This time point, which approximately coincides with the onset of the YFP "death oscillation", serves as our defined boundary between the "alive" and "dying" phases of cellular tracks. Therefore, for all further analyses of clock characteristics, we will relate only to "alive" tracks which we define by those that are free of "death oscillation" for at least 60 h after the cold temperature shift.

## 3. The transitory dynamics of the circadian clock is well described by the Stuart-Landau model

Figures 2 a, b, and c show a few randomly chosen individual among "alive" tracks of YFP from 3 different experiments. Each track is monitored for at least 2.5 days at 25°C before being switched to colder temperatures. A sinusoidal fitting, limited to a short two-day period (lines in Fig. 2a, b, c) allows to extract the individual period and amplitude for each fitted track at both temperatures (dots respectively in Fig. 2d, e). One can see that, when the temperature is lowered, the circadian period only slightly increases (Fig. 2d) while the average amplitude strongly decreases (Fig. 2e). Additionally, as observed in Figure 2 b and c, most tracks show a decreasing oscillation amplitude over time and this decrease occurs tremendously faster for lower temperatures.

To gain insights into the dynamical aspect of the oscillator, we replace a direct observable, the density of fluorescence, with the reporter promoter activity (PA). The promoter here is PkaiBC, a class I promoter, regulated by the phosphorylation level of KaiC whose activity directly relates to the dynamical state of the *in vivo* circadian oscillator [18, 19]. Figures 2 f, g, and h show an example of the PA (open circles) for two representative tracks for each condition selected to be longer than 96 h at cold temperatures. This confirms that even after 4 days in any of the low temperature conditions, the oscillation track is not at steady state, as the amplitude continues to decrease over time.

To overcome this issue, and capture both amplitude and phase dynamics across time and temperature, we use the Stuart-Landau formalism [20, 21] which provides a rigorous mathematical framework for analyzing nonlinear oscillators near Hopf bifurcations[22]. The Stuart-Landau model tracks the instantaneous phase $\theta(t)$ and amplitude $r(t)$ following the dynamical equation of the complex variable $\underline{z} = r \cdot e^{i\theta}$ (represented zero centered):

$$\frac{d\underline{z}}{dt} = (\mu + i\omega_c)\underline{z} - (a - ib)\underline{z}|\underline{z}|^2 \quad (1a)$$

or equivalently in polar coordinates:

$$\begin{cases} \dot{r} = \mu r - ar^3 \\ \dot{\theta} = \omega_c + br^2 \end{cases} \quad (1b)$$

In our analysis, the amplitude $r(t)$ is already normalized by the initial value of the amplitude (Supplementary Information), considered here at steady-state. The assumption of initial steady state is reasonable as it takes approximately one day for the circadian clock in individual cells[6, 8] to adapt to the change in their surroundings (transfer from a flask in the incubator at 25°C to a microscope observation chamber incubated at the same temperature). With this assumption, the parameters $a, b$ are set by the initial steady state conditions (here $T_0 = 25°C$), while the parameters $\mu(T)$ and $\omega_c(T)$ depend on the final temperature (here the cold temperature $T$). Note that $a, b, \omega_c$ here are strictly positive, while $\mu$, the bifurcation control parameter, can have both signs.

For $\mu > 0$, after a transitory time, a stable oscillation of amplitude $r^* = \sqrt{\mu/a}$ and angular frequency $\omega^* = \omega_c + b\mu/a$ is reached. An example of this transition is given in Fig. 2f and i, after temperature is switched form 25°C to 21°C: the oscillation quits the larger limit cycle, of amplitude 1, to reach, after a couple of oscillations, a lower amplitude orbit. Additionally, the initial steady state coordinates expression, $r_0^* = 1$, $\omega_0^* = \omega_c(T_0) + b$, allows us to determine $a = \mu(T_0) = \mu_0$ and $b = \omega_0^* - \omega_c(T_0)$.

For the case $\mu < 0$, the trajectory converges toward the fixed point (the origin) with a limit angular frequency $\omega_c$, and amplitude $r^* = 0$ (Fig. 2 h and k). The transition between the two regimes is a textbook [23] presentation of the supercritical Hopf bifurcation with the critical control parameter $\mu = 0$, and angular frequency $\omega_c$.

Regardless of the sign of $\mu$, the time scale for the oscillator to reach the new steady state is mainly given by $|\mu|^{-1}$: the closer is the system to the bifurcation, the longer it takes to reach the new steady state. The example in the Fig. 2 g and j is an illustration of a long transitory time, where the oscillator orbits the origin numerous rounds before reaching the final steady-state close to the fixed point.

The system (1) has an analytic solution (Supplementary Information) and we can fit the transitory part of the warm to cold transition, by $y_{cold}(t) = r(t) \cdot cos(\varphi(t)) + baseline_{cold}$, where $r(t)$ and $\varphi(t)$ are extracted from:

$$\begin{cases} r^2(t) = \dfrac{1}{\dfrac{\mu_0}{\mu} + \left(1 - \dfrac{\mu_0}{\mu}\right)e^{-2\mu t}} \\ \varphi(t) = \varphi_{ini} + \omega_c t + \dfrac{b}{2\mu_0} \cdot \ln\left[\dfrac{\mu_0}{\mu}(e^{2\mu t} - 1) + 1\right] \end{cases} \quad (2)$$

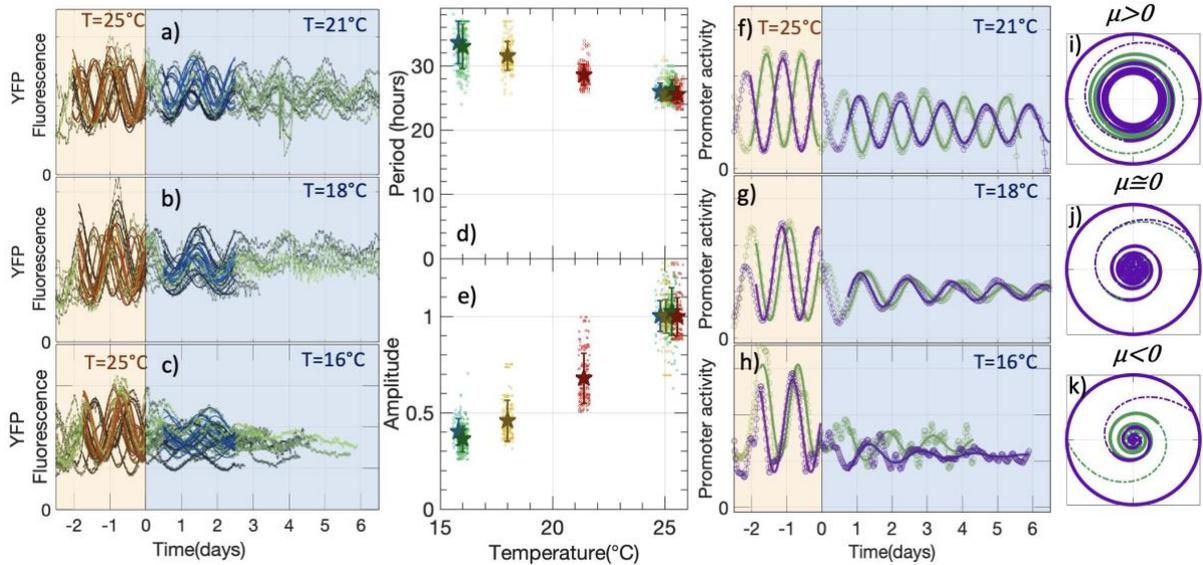

*Figure 2 The in vivo oscillation transitory track described by a Stuart Landau oscillator a-c) Density of YFP fluorescence/ cell, for 3 cold temperatures (21°C, 18°C and 16°C) over time. Time, in days (1 day = 24-hour) starts at the cold onset. Approximately 10% of the total fitted tracks, randomly chosen are shown (open circles, in shades of green). The results of the fitting by the function, $y(t) = A \cdot cos(2\pi t/T + \varphi_0) + baseline$, are represented by the continuous lines in orange shade (at 25°C) or blue shades (at colder temperatures). d) The resulting period and e) amplitude as a function of temperature (colors represent experiments). Each individual dots are from the fit of one track (at 25°C and at the respective cold temperature). Individual fitting error bars are omitted for clarity but used for mean (☆) and s.e.m of the entire set data of whole experiment. For better visibility individual point are spread-out $\cong 0.6°C$ on x-axis, comparable to the uncertainty of the temperature. The warm temperature results (25°C) are slightly shifted for distinction. The represented amplitude in d) is the normalized value (by the average of the progenitor colony at 25°C). f-h) Two representative normalized long tracks (green and purple circles) of kaiBC PA for each cold temperature (21°C, 18°C and 16°C) as a function of time. The continuous lines, in the warm part represent a fit with the function $y_{warm}(t) = cos(2\pi\omega_0 t + \varphi_0) + baseline_{warm}$ in the time interval $(-1.7 < t \leq 0)$. The continuous line in the cold part (blue background) is a SL fit (Eq 2) starting at $t > 0.7$ days. i-k) reconstituted trajectories of the oscillators from figure (f-h): the larger cycle corresponds to the initial steady-state warm fit of amplitude 1, then the transitory part of the SL fit extended to $(0 < t \leq 20)$.*

For higher precision, we take into consideration the cell divisions and fit a *whole* lineage at once. As cells continue to divide after the cold onset, a progenitor before the cold transition will divide, and the track will separate in several different tracks with a shared initial condition. The resulting fit has 6 parameters ($b, \varphi_{ini}, \mu_0, \omega_c, \mu, baseline_{cold}$), 3 of them ($b, \varphi_{ini}, \mu_0$) are shared within the lineage, while the 3 others ($\omega_c, \mu, baseline_{cold}$) are specific to individual tracks (see Supplementary Information, Fig. S2). An example of fitting for two independent tracks is represented by the continuous lines in Figures 2f, g, and h. In the warm temperature region, the fit is a steady-state oscillation, therefore a sinusoidal curve like $y_{warm}(t) = cos(2\pi\omega_0 t + \varphi_0) + baseline_{warm}$. In the cold region, the transitory response of the oscillator is well described by the Stuart-Landau formalism (Eq 2).

We performed the same analysis for all the tracks in our experiments and Figures 3 a, b show the temperature dependence of the two main parameters $\omega_c(T)$ and $\mu(T)$. Note that, with our assumption of initial steady state we also obtain an estimation of these parameters at 25°C. As the temperature lowers, the average control parameter of the bifurcation $\langle \mu \rangle$, changes its sign, confirming therefore that the *in vivo* circadian oscillator indeed goes through a Hopf bifurcation. The temperature at which the bifurcation happens, extracted by a linear fit of $\langle \mu(T) \rangle$ is $T_{in\,vivo} = 19°C \pm 1°C$ (green shade interval in Fig 3b).

We applied a similar Stuart-Landau formalism (Supplementary Information) to the *in vitro* tracks results from [3] and fitted each individual transitory track. The fit shown in Fig. S3, and the extracted parameters $\omega_{c\,in\,vitro}(T)$ and $\mu_{in\,vitro}(T)$ are plotted for comparison on the same Fig. 3 a, b. The temperature dependence of the control parameter $\mu_{in\,vitro}$ provides (trough a similar linear fit) an estimation of the bifurcation temperature $Tc_{in\,vitro} = 21.4°C \pm 0.6°C$ (at 95% confidence interval – blue shade interval). We have therefore both circadian oscillators, *in vivo* and *in vitro*, in a close temperature range well described by a Stuart-Landau model. In the last part of this work, we examine how these two temperature dependent descriptions are related.

## 4. Two scenarios of embedded Stuart-Landau with feedback explain the same set of data

Intensive studies of the circadian clockwork in *Synechococcus elongatus* have uncovered several embedded cycles (Fig. 3c). At the heart of the clock is the *in vitro* oscillator, driven only by proteins and ATP interactions. This mass-action biochemical oscillator consists of two intertwined cycles: the KaiABC nanocomplex phosphorylation cycle and the KaiC-ATPase cycle. The complete *in vitro* oscillator drives the *in vivo* oscillator, along with many layers of regulation [24], summed up in a complementary cycle generated by a transcription translation feedback loop [25, 26, 7].

Several analytical and numerical models have been developed to describe both oscillators and their interactions [27, 28]. However, few models account for temperature dependence [29, 30, 31, 32] none address the *in vivo* system, and none focus on the crucial transitory features described above. Our aim is to develop the simplest feedback mechanism that will remap the *in vitro* temperature variation of $\omega_{c\,in\,vitro}(T)$ and $\mu_{in\,vitro}(T)$ to the *in vivo* one $\omega_{c\,in\,vivo}(T)$ and $\mu_{in\,vitro}(T)$.

## 4.1 Instantaneous feedback

The simplest extension of the model used above is a Stuart-Landau oscillator with an instantaneous feedback term added to Eq (1b):

$$\frac{dz}{dt} = (\mu + i\omega_c)\underline{z} - (a - ib)\underline{z} \mid z \mid^2 + Ke^{i\theta}\underline{z} \tag{3}$$

$K$ being the feedback strength and $\theta$ the coupling phase. The coupling strength $K$ has the same dimension as $\mu$ or $\omega_c$, therefore it directly compares to them, and measures the proportion of the output signal that is feedbacked. The phase $\theta$, on the other hand, captures how this feedback is shared between the two dimensions of the limit cycle.

If the *in vitro* oscillator were a SL oscillator of parameters $\omega_c = \omega_{c\,in\,vitro}(T)$ and $\mu = \mu_{in\,vitro}(T)$, the result of the feedback would again be a SL oscillator with the effective new parameters (Supplementary Information) $\omega_{c_{vivo}} = \omega_c + K\sin\theta$ and $\mu_{vivo} = \mu + K\cos\theta$. Figures 3 d, e show (in blue shades) the resulting coupling strength $K(T)$ and phase $\theta(T)$. The resulting feedback constant, always positive ($K(T) > 0$), is weaker close to the bifurcation temperature and stronger far away. The phase of the feedback, $\theta$, changes its sign with temperature within a limited interval $-\frac{\pi}{2} < \theta(T) < \frac{\pi}{2}$ which results in $K\cos\theta(T) > 0$.

## 4.2 Delayed feedback

A second simple and plausible feedback model includes a time delay in the feedback loop, reflecting the inherent delays in many biochemical reaction networks due to the sequential assembly of functional proteins through transcription, translation, folding, and phosphorylation. In this model, the output of the *in vitro* oscillator requires a specific duration, denoted as $\tau$, before it feeds back into the system with a strength $K$:

$$\frac{dz}{dt} = (\mu + i\omega_c)\underline{z} - (a - ib)\underline{z} \mid z \mid^2 + K\underline{z}(t - \tau) \tag{4}$$

Delayed feedback oscillators are well-known in various fields of dynamics [33] including optical systems, control systems, economic cycles, population dynamics, neural networks, and gene regulation. Depending on the parameter values, Equation (4) can yield diverse solutions, from steady-state oscillations to extinction of initial oscillations, multi-stable states, and even chaos[34]. The experimental constrains of our experiments allow us to limit the possible outcomes to two: a steady-state oscillation or amplitude extinction (Supplementary Information).

For delayed feedback, the transitory regime cannot be solved analytically and must be addressed numerically to extract $K$ and $\tau$ parameters as detailed in Supplementary Information. The procedure involves few steps: we first show that the apparent *in vivo* SL parameters ($\mu_{vivo}, \omega_{c_{vivo}}$) as a function of ($\mu_{vitro}, \omega_{c_{vitro}}$) are solution of the transcendent equations (eq S-25, S-28). Next, in Supplementary Information, we illustrate that the parameters inferred by SL fit of the transitory track correspond reasonably well with the analytic solution in both regimes. In the last step of the procedure, we use a non-linear fit in order to obtain the parameters ($K, \tau$) that optimize the remapping of ($\mu_{vitro}, \omega_{c_{vitro}}$) to ($\mu_{vivo}, \omega_{c_{vivo}}$) trough eq S-25, S-28.

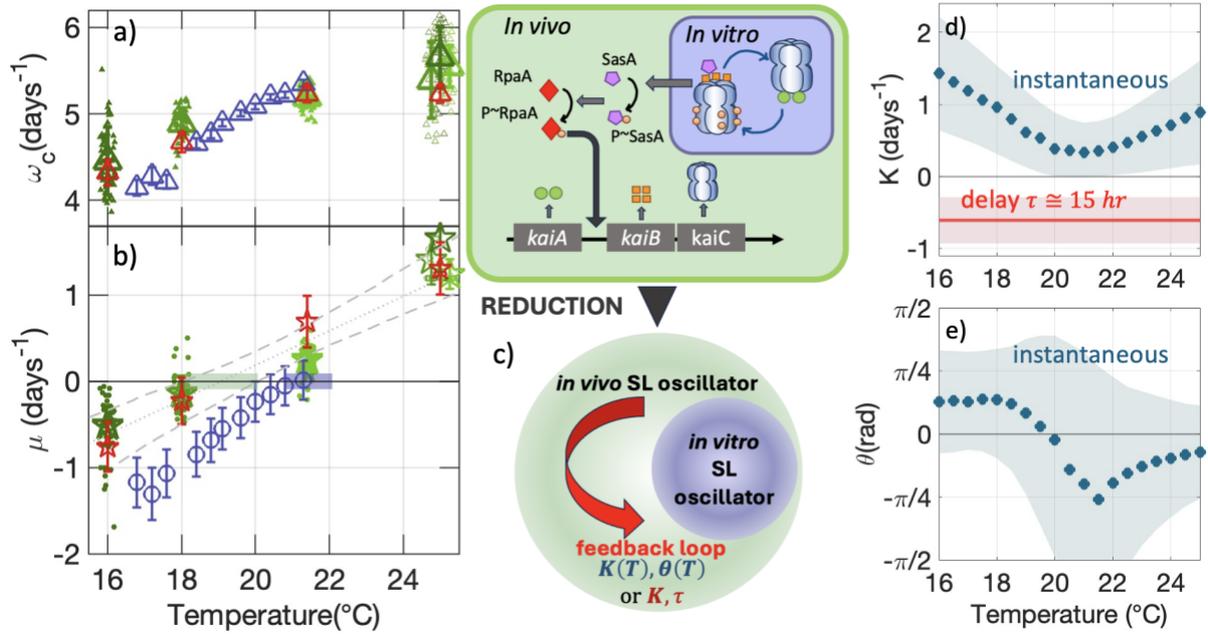

*Figure 3 The in vivo and in vitro oscillators go through a similar Hopf bifurcation and feedback can transform the in vitro system into the in vivo one. a-b) A synthesis of fit results, the average and s.e.m $\omega_c$ (Δ) and μ (☆) versus temperature (16°C, to 25°C, dark green to light green). Individual track fits shown as small dots, spread ≅0.6°C on x-axis for visibility. All the warm temperature results (25°C) are slightly shifted to be individually discernible. In vitro P~KaiC track fits in dark blue with the fitting error bar, $\omega_c$ (Δ) and μ (○). The dashed lines in (b) are the 95% confidence interval of a linear fit of μ as a function of temperature. The two horizontal filled bars, represent the 95% confidence interval for the critical temperature (where μ=0), in green, the in vivo value, in dark purple the in vitro case. c) Upper panel: a representation simplified of the molecular model of the interaction between the cyanobacterial in vitro clock and the in vivo one. Lower panel: the reduction considered here. An in vitro Stuart-Landau oscillator (the dark blue disk) with a simple feedback loop to become the in vivo one (green disk). In the first scenario, the feedback is instantaneous with the coupling constant K(T) and the phase θ(T) temperature dependent (blue curves in d-e, with the clearer shade the 95% confidence intervals). In the second scenario the feedback is delayed by τ before being applied with a coupling constant K. In this last scenario both τ and K are temperature independent (red line in d, with the clearer shade the 95% confidence interval).*

The final result is surprisingly simple: a single set of constant values $(K, \tau)$ suffices to describe the relation between the whole set of *in vitro* SL parameters $(\mu_{vitro}(T), \omega_{c_{vitro}}(T))$ and the *in vivo* ones $(\mu_{vivo}(T), \omega_{c_{vivo}}(T))$. The value for the amplitude coupling is $K = -0.61 \pm 0.32 \text{ days}^{-1}$ within a 95% confidence interval (represented in a red line in Fig 3d) and indicating that the feedback remains constant within the studied temperature range. This negative feedback is of mild strength compared to the overall variations of μ or $\omega_c$. The delay is $\tau = 15.4 \pm 2$ h, also constant in temperature.

In essence, we have two different interpretations of the same dataset. The first scenario proposes immediate positive feedback that varies with temperature, while the second suggests time delayed negative feedback that is temperature compensated. While both views are admittedly oversimplified, they still prompt us to discuss what the dominant feedback for the *in vivo* clock might be and how it depends on temperature.

## 5. Discussion

A dynamical system, such as a limit cycle oscillator, reveals itself through perturbations. The collective efforts of the last decades have uncovered the general and detailed clockwork of both *in vitro* and *in vivo* systems in cyanobacteria, through molecular or environmental disruptions of the underlining network. The novelty in this article is that we quantitatively analyze the transitory phase in between two steady states to extract the information embedded within the clock's complex network and the link between both oscillators, the *in vivo* and the *in vitro* ones.

We first explored *in vivo* imaging of the clock at lower temperatures. Few studies have examined low-temperature effects on single bacteria and none yet using time-lapse microscopy. We found that *Synechococcus elongatus* PCC7942 has a significantly reduced survival rate under chilling temperatures when monitored this way. The strain's sensitivity to cold was unexpected, given its origin [12] in San Francisco Bay spring water. This could be due to the microscopy environment's light flashes (for fluorescence monitoring) or the strain's adaptation to optimal lab conditions (30°C) over 40 years, potentially losing cold tolerance mechanisms. Future research could explore these lost adaptations and examine the clock at lower temperatures. Nevertheless, we identified a sequence of death steps based on growth rate, thylakoid membrane fluorescence, and *PkaiBC:yfp* reporter fluorescence. Based on that we were able to untangle this sequence from the circadian response and focus solely on the survivors.

By tracking only "alive" cells, we found that the *in vivo* circadian oscillator undergoes a supercritical Hopf bifurcation. We quantified this bifurcation by fitting the transitory regime with a Stuart-Landau model, obtaining the two parameters describing the bifurcation: $(\mu, \omega_c)$ for each track at the low temperature and $(\mu_0, \omega_{c_0})$, characteristic of the common ancestor bacterial cell at the initial temperature. Similarly, we reinterpreted from[3], the transitory oscillation of the *in vitro* P~KaiC (proportion of phosphorylated KaiC) curves and extracted the temperature dependence of the parameters $(\mu_{vitro}(T), \omega_{c\,vitro}(T))$.

The first observation that stands out is the similarity in bifurcation temperatures between the two oscillators. As the temperature decreases, the *in vitro* clock is the first to approach its extinction point, closely followed by the *in vivo* oscillator. This is not entirely surprising, as the *in vivo* clock is more stable and robust against perturbations [7]. The second observation is that $\mu(T)$ and $\omega_c(T)$ for *in vivo* and *in vitro* vary respectively over similar range of values, with only slightly different slopes.

A comprehensive theoretical description, the Sasai model[30], based on numerical simulations coupled with thermodynamic aspects, takes into account the biochemical and structural features of the *in vitro* clock. Using this model, we simulated *in vitro* oscillator transitory tracks (Supplementary Information) and then applied the same fitting protocol as for the experimental results. The Sasai oscillator undergoes a Hopf bifurcation with comparable values for the bifurcation temperature and for $\omega_c(T)$, but the $\mu(T)$ curve is nearly one order of magnitude lower (Fig. S11) than the experimental one. Another recent theoretical work [32], introduces a new analytical perspective that explains the temperature compensation in the cyanobacterial oscillator through oscillation-phase separation: the oscillation is supposed to be far from a sinusoidal-like cycle and thus far from the Hopf bifurcation. However, as we demonstrate here, this contradicts our experimental results. Therefore, the results presented in this paper, provide strong constraints, and similar time-resolved analysis need to be considered in further models.

In the final part of this work, we demonstrate that an in *vivo* SL oscillator with feedback can explain at once both *in vivo* and *in vitro* temperature dependence of $\mu$ and $\omega_c$. Two opposing scenarios of feedback can explain this behavior: either instantaneous, temperature-dependent positive feedback or time delayed negative feedback (Fig. 3 d, e). It is worth noting that in the second scenario, both the time delay and the feedback coupling are constant with respect to temperature.

The cycle of phosphorylation/dephosphorylation of KaiC within the KaiC hexamer drives the *in vitro* clock, while only the active fold switched fsKaiB [35] can bind to the KaiC heterocomplex and then sequester KaiA for global hexamer clock synchronization [36]. According to previous studies [7, 37], the main *in vivo* feedback loop is negative and it concerns SasA [38] : while associated with the Kai complex, SasA phosphorylate the master regulator RpaA to P~RpaA. In turn, P~RpaA globally regulates transcription [39] in this bacterium and more specifically activates the transcription of the class I promoters, like PkaiBC. The newly produced ground state fold of KaiB (gsKaiB) and unphosphorylated KaiC monomers from their shared promoter diminish the proportion of phosphorylated KaiC(P~KaiC) and of the active fold switched KaiB (fsKaiB), therefore acts as effective negative feedback.

The *in vivo* circadian clock in cyanobacteria is also known for its robustness against perturbations and stochastic noise at the individual cell level [6]. The transcription-translation regulation is responsible for this resilience [7]. A negative feedback loop with noisy distributed delay has recently been demonstrated to sharpen oscillation peaks and improve temporal reliability without affecting the period [40]. We expect to have a noisy delay at every step along the feedback loop, as transcription, translation, folding, and post-transcriptional processes are all subject to stochastic noise and may contribute to a global distributed noisy delay. Consequently, these observations further support the presence of a dominant delayed negative feedback loop.

That said, at the molecular level, many additional feedback loops come into play *in vivo*. These include the rhythmic production of partner proteins such as SasA, CikA, and KidA [41]; the rhythmic degradation of Kai proteins; and the recently reported complementary function of SasA as a hetero-cooperative enhancer of KaiB's association with KaiC [42]. In this context, although we favor the time-delayed negative feedback model we cannot be certain of its relevence. This uncertainty is partly because the SL *in vitro* description may not have a direct equivalent to simple measured quantities.

Further theoretical and experimental work, similar to that on the glycolytic oscillator [43, 44] should be conducted to understand and reduce the complex biochemical map of the *in vitro* oscillator near the bifurcation point. While numerous experiments exist in this domain, few focus on the transitory regime and even fewer have done that at different temperatures. Recent experiments setup that reconstitutes the *in vitro* oscillator and *in vivo*-like clock in an optical microplate reader [42] might be, the next step to test the clock's response to different temperature steps.

**Author contributions:** Conceptualization: I.M., H.I and B.H., Methodology: M.V-G and J.G Investigation: I.M. and H.I. Image analysis curation: M.H; K.H, I.M and I.H. Formal analysis and modeling: I.M., B.H. and H.I., Writing – original draft: I.M. Writing – review and editing: I.M., B.H. and H.I. Funding acquisition: I.M and H.I

**Competing** interests: The authors declare no competing interests.

**Data and materials availability:** All data are available in the main text or the supplementary materials.

## Material and Methods

### Strains & culture

Wild type *Synechococcus elongatus* PCC 7942 as well as the ΔkaiBC background, carrying P*kaiBC-yfp-lvA* reporter in NSI, respectively JRCS32 and JRCS65 [18] were precultured in modified BG-11 liquid medium under continuous light (LL) of 40 µmol·m$^{-2}$·s$^{-1}$ at 25 °C.

### Sample preparation

We modified the protocols proposed by [45]. To 3 ml of 2x BG11 medium in a 50-ml polypropylene conical tube kept at 30 ºC, we added 3 ml of 3% (wt/vol) low-melt agarose dissolved by microwaving and vortexed thoroughly. Then we add 500 µl of the agarose mixture on the cover glass slide and immediately put another cover glass to create an agarose pad. After drying the pad for 25 mins at room temperature, we removed a cover glass of one side and placed several drops containing bacteria (OD ~ 0.05) on the pad. We cut out a smaller pad containing the drop (~3mm x 3mm) with a scalpel and then flipped it onto a cover glass-bottom dish (Mattek). We poured 150 µl of the mixture of melted agarose and BG11 into the bottom of the dish, which was prepared at the beginning of this experiment. The agarose with the pad was covered with a 0.4µm pore size, transparent PET membrane filter (Falcon, 353090) and sealed with low melting paraffin wax (Fluka, 76243). The upper space, above the filter, was filled with liquid BG11 medium. We made a 2mm hole on the petri-dish cover to allow the *in situ* temperature sensor to be bathed in this BG11 medium, in close vicinity to the cyanobacteria microscopic location.

### Microscopy

Time lapse images of fluorescence and phase contrast were taken every 30 min, over 5-10 fields of each of the 2 strains (wild-type kaiC-promoter and delta) for experiments as long as 10 days. We used for that an inverted AxioObserver7 Zeiss microscope with a Plan-Apochromat x63 NA 1.4, Ph3 oil objective and an automated autofocus (Definite Focus2). For the fluorescence monitoring a LED setup (Zeiss Colibri7) was used as follows: for the yfp fluorescence, the 515nm line was associated with Chroma 49003 filter set and for the red autofluorescence the 555nm line associated with the Chroma 49005 filter set. All images were acquired using the Hamamatsu - Orca FlashV3 CMOS camera. The automatization of shutters, XY-stage movement, autofocus and image acquisition was done via Micromanager [46] associated with a custom routine for controlling a ring of white LED (cells were illuminated at 2500 lux except during image acquisition). Temperature of the sample was set via a custom sample holder temperature controlled by a water bath (Lauda RE104). In a second set of experiments, the whole microscope was surrounded by a custom box, cooled/heated by a Peltier element with PID controller (TE Technologies). Heating up and cooling down the sample took approximately 30 min. The *in-situ* temperature of the sample as well the light received by bacteria was continuously monitored by Vernier probes (STS-BTA, LS-BTA).

### Image and data analysis

Phase contrast and fluorescence images were assembled and registered using ImageJ [47]. The hyper-stacks (time, phase, fluorescence channels) of individual microcolonies were first automatically segmented and linked using SuperSegger [48] using a custom trained model and then manually curated for the remaining segmentation and linking errors. Individual and lineage cell characteristics were then extracted and analyzed using Matlab custom routines.

### Estimation of the promoter activity from time-lapse data

For the instantaneous promoter activity *PA(t)* estimate we used the expression: $PA(t) = [dil(t) + \gamma(t)] \cdot M(t) + \frac{dM(t)}{dt}$, with *M(t)* the instantaneous density of fluorescence of a cell, *dil(t)* the instantaneous growth rate and $\gamma(t)$ the overall degradation constant of fluorophores including both kind, active degradation of fluorescent reporter $\gamma_{deg}$ and bleaching $\gamma_{bleach}$: $\gamma(t) = \gamma_{deg}(t) + \gamma_{bleach}$. The density of fluorescence at a given time point $t_i$, $M(t_i)$, is defined the sum of all fluorescence values for each pixel belonging to a cell $f_j(t_i)$ at the time $t_i$ divided by the number of pixels of the cell, minus the microcolony local value of the fluorescence background $bg(t_i)$ : $M(t_i) = \frac{\sum_{j=1}^{Npixels(t_i)} f_j(t_i)}{Npixels(t_i)} - bg(t_i)$. For PA calculation, *M(t$_i$)* has been smoothed with a local regression method ('loess" smooth with a span of 10% in Matlab) and then used to numerically take the derivative. The instantaneous growth rate at a given time point $t_i$, $dil(t_i)$, was estimated by using the 50% "loess" smooth of individual cell length track $L(t_i)$, then numerically derived to obtain : $dil(t_i) = \frac{1}{L(t_i)} \cdot \frac{dL}{dt}\big|_{t_i}$. The bleaching degradation constant $\gamma_{bleach}$ is dependent mainly on the light exposure during image acquisition and was considered constant throughout the experiment. The active degradation constant of the LVA-SsrA-tagged reporter protein, $\gamma_{deg}$ is temperature dependent [49] therefore will vary along the experiment and from one experiment to another, when switching from warm to cold regime. However, we show that the effect on the fitted oscillator parameters extraction is minimal Supplementary Information.

**Fit and statistics**

Single tracks of "alive" cell were selected as explained in the main text. For PA activity fitting with Stuart-Landau equation, only the long tracks were selected, meaning "alive" for longer that 96 h after the cold onset. While in the cold regime the cells are less dividing, duplicate statistics on the warm region have been excluded by resorting the tracks by families: all the tracks which are identical within last day previous the cold downshift will belong to the same family and the warm track is the progenitor. All statistics in the warm region are done only on the progenitor. The individual tracks fit was done by using the Matlab nonlinear fitting routines as explained in the main text and in Supplementary Information. Note that the Stuart-Landau fitting is done for a given family considering common initial conditions. In the end a bootstrap method has been next used to obtain the statistics on the parameters along with their individual confidence interval.

# 1.1 Bibliography

1. Njus, D., McMurry, L. & Hastings, J. W. Conditionality of circadian rhythmicity: Synergistic action of light and temperature. *J. Comp. Physiol. B* **117**, 335–344 (1977).
2. Revel, F. G. *et al.* The circadian clock stops ticking during deep hibernation in the European hamster. *Proc. Natl. Acad. Sci.* **104**, 13816–13820 (2007).
3. Murayama, Y. *et al.* Low temperature nullifies the circadian clock in cyanobacteria through Hopf bifurcation. *Proc. Natl. Acad. Sci.* **114**, 5641–5646 (2017).
4. Enoki, R. *et al.* Cold-induced suspension and resetting of Ca2+ and transcriptional rhythms in the suprachiasmatic nucleus neurons. *iScience* **26**, 108390 (2023).
5. NAKAJIMA, M. *et al.* Reconstitution of circadian oscillation of cyanobacterial KaiC phosphorylation in vitro. *Reconst. Circadian Oscil. Cyanobacterial KaiC Phosphorylation Vitro* **308**, 414–415 (2005).
6. Mihalcescu, I., Hsing, W. & Leibler, S. Resilient circadian oscillator revealed in individual cyanobacteria. *Nature* **430**, 81–85 (2004).
7. Teng, S.-W., Mukherji, S., Moffitt, J. R., de Buyl, S. & O'Shea, E. K. Robust Circadian Oscillations in Growing Cyanobacteria Require Transcriptional Feedback. *Science* **340**, 737–740 (2013).
8. Gan, S. & O'Shea, E. K. An Unstable Singularity Underlies Stochastic Phasing of the Circadian Clock in Individual Cyanobacterial Cells. *Mol. Cell* **67**, 659-672.e12 (2017).
9. Zakhia, F., Jungblut, A., Taton, A., Vincent, W. & Wilmotte, A. Cyanobacteria in Cold Ecosystems. in *Psychrophiles: From Biodiversity to Biotechnology* 121–135 (2008). doi:10.1007/978-3-540-74335-4_8.
10. Los, D. & Murata, N. Sensing and Responses to Low Temperature in Cyanobacteria. in *Cell and Molecular Response to Stress* vol. 3 139–153 (2002).
11. Sinetova, M. A. & Los, D. A. New insights in cyanobacterial cold stress responses: Genes, sensors, and molecular triggers. *Biochim. Biophys. Acta BBA - Gen. Subj.* **1860**, 2391–2403 (2016).
12. Golden, S. S. The international journeys and aliases of Synechococcus elongatus. *N. Z. J. Bot.* **57**, 70–75 (2019).
13. Ohbayashi, R. *et al.* Coordination of Polyploid Chromosome Replication with Cell Size and Growth in a Cyanobacterium. *mBio* **10**, 10.1128/mbio.00510-19 (2019).
14. Yu, J. *et al.* Synechococcus elongatus UTEX 2973, a fast growing cyanobacterial chassis for biosynthesis using light and CO2. *Sci. Rep.* **5**, 8132 (2015).
16. Porankiewicz, J. & Clarke, A. K. Induction of the heat shock protein ClpB affects cold acclimation in the cyanobacterium Synechococcus sp. strain PCC 7942. *J. Bacteriol.* **179**, 5111–5117 (1997).
17. Material and methods.
18. Chabot, J. R., Pedraza, J. M., Luitel, P. & van Oudenaarden, A. Stochastic gene expression out-of-steady-state in the cyanobacterial circadian clock. *Nature* **450**, 1249–1252 (2007).
19. Murayama, Y., Oyama, T. & Kondo, T. Regulation of Circadian Clock Gene Expression by Phosphorylation States of KaiC in Cyanobacteria. *J. Bacteriol.* **190**, 1691–1698 (2008).
20. A. A. Andronow & C. E. Chaikin. *Theory Of Oscillations*. (1949).
21. García-Morales, V. & Krischer, K. The complex Ginzburg–Landau equation: an introduction. *Contemp. Phys.* **53**, 79–95 (2012).
22. Kuramoto, Y. *Chemical Oscillations, Waves, and Turbulence*. (Courier Corporation, 2003).
23. Strogatz, S. *Nonlinear Dynamics and Chaos: With Applications to Physics, Biology, Chemistry, and Engineering (Studies in Nonlinearity)*. (Westview Press, 2000).
24. Cohen, S. E. & Golden, S. S. Circadian Rhythms in Cyanobacteria. *Microbiol. Mol. Biol. Rev.* **79**, 373–385 (2015).
25. Kitayama, Y., Nishiwaki, T., Terauchi, K. & Kondo, T. Dual KaiC-based oscillations constitute the circadian system of cyanobacteria. *Genes Dev.* **22**, 000–000 (2008).
26. Qin, X., Byrne, M., Xu, Y., Mori, T. & Johnson, C. H. Coupling of a Core Post-Translational Pacemaker to a Slave Transcription/Translation Feedback Loop in a Circadian System. *PLOS Biol.* **8**, e1000394 (2010).

27. van Zon, J. S., Lubensky, D. K., Altena, P. R. H. & ten Wolde, P. R. An allosteric model of circadian KaiC phosphorylation. *Proc. Natl. Acad. Sci.* **104**, 7420–7425 (2007).
28. Rust, M. J., Markson, J. S., Lane, W. S., Fisher, D. S. & O'Shea, E. K. Ordered Phosphorylation Governs Oscillation of a Three-Protein Circadian Clock. *Science* **318**, 809–812 (2007).
29. Hatakeyama, T. S. & Kaneko, K. Generic temperature compensation of biological clocks by autonomous regulation of catalyst concentration. *Proc. Natl. Acad. Sci.* **109**, 8109–8114 (2012).
30. Sasai, M. Role of the reaction-structure coupling in temperature compensation of the KaiABC circadian rhythm. *PLOS Comput. Biol.* **18**, e1010494 (2022).
31. Chakravarty, S., Hong, C. I. & Csikász-Nagy, A. Systematic analysis of negative and positive feedback loops for robustness and temperature compensation in circadian rhythms. *NPJ Syst. Biol. Appl.* **9**, 5 (2023).
32. Fu, H., Fei, C., Ouyang, Q. & Tu, Y. Temperature compensation through kinetic regulation in biochemical oscillators. *Proc. Natl. Acad. Sci.* **121**, e2401567121 (2024).
33. *Applied Delay Differential Equations*. (Springer, New York, NY, 2009). doi:10.1007/978-0-387-74372-1.
34. D'Huys, O., Vicente, R., Danckaert, J. & Fischer, I. Amplitude and phase effects on the synchronization of delay-coupled oscillators. *Chaos Interdiscip. J. Nonlinear Sci.* **20**, 043127 (2010).
35. Swan, J. A. *et al.* Coupling of distant ATPase domains in the circadian clock protein KaiC. *Nat. Struct. Mol. Biol.* **29**, 759–766 (2022).
36. Sasai, M. Mechanism of autonomous synchronization of the circadian KaiABC rhythm. *Sci. Rep.* **11**, 4713 (2021).
37. Nakahira, Y. *et al.* Global gene repression by KaiC as a master process of prokaryotic circadian system. *Proc. Natl. Acad. Sci.* **101**, 881–885 (2004).
38. Takai, N. *et al.* A KaiC-associating SasA–RpaA two-component regulatory system as a major circadian timing mediator in cyanobacteria. *Proc. Natl. Acad. Sci.* **103**, 12109–12114 (2006).
39. Markson, J. S., Piechura, J. R., Puszynska, A. M. & O'Shea, E. K. Circadian control of global gene expression by the cyanobacterial master regulator RpaA. *Cell* **155**, 1396–1408 (2013).
40. Song, Y. M., Campbell, S., Shiau, L., Kim, J. K. & Ott, W. Noisy Delay Denoises Biochemical Oscillators. *Phys. Rev. Lett.* **132**, 078402 (2024).
41. Kim, S. J., Chi, C., Pattanayak, G., Dinner, A. R. & Rust, M. J. KidA, a multi-PAS domain protein, tunes the period of the cyanobacterial circadian oscillator. *Proc. Natl. Acad. Sci.* **119**, e2202426119 (2022).
42. Chavan, A. G. *et al.* Reconstitution of an intact clock reveals mechanisms of circadian timekeeping. *Science* **374**, eabd4453 (2021).
43. Hynne, F., Danø, S. & Sørensen, P. G. Full-scale model of glycolysis in *Saccharomyces cerevisiae*. *Biophys. Chem.* **94**, 121–163 (2001).
44. Danø, S., Madsen, M. F., Schmidt, H. & Cedersund, G. Reduction of a biochemical model with preservation of its basic dynamic properties. *FEBS J.* **273**, 4862–4877 (2006).
45. Young, J. W. *et al.* Measuring single-cell gene expression dynamics in bacteria using fluorescence time-lapse microscopy. *Nat. Protoc.* **7**, 80–88 (2012).
46. Edelstein, A. D. *et al.* Advanced methods of microscope control using μManager software. *J. Biol. Methods* **1**, e10 (2014).
47. Schindelin, J. *et al.* Fiji: an open-source platform for biological-image analysis. *Nat. Methods* **9**, 676–682 (2012).
48. Stylianidou, S., Brennan, C., Nissen, S. B., Kuwada, N. J. & Wiggins, P. A. SuperSegger: robust image segmentation, analysis and lineage tracking of bacterial cells. *Mol. Microbiol.* **102**, 690–700 (2016).
49. Cordova, J. C. *et al.* Stochastic but Highly Coordinated Protein Unfolding and Translocation by the ClpXP Proteolytic Machine. *Cell* **158**, 647–658 (2014).